\documentclass[a4paper]{jpconf}
\usepackage{graphicx}
\usepackage{amsmath,amssymb}
\usepackage{epsfig}
\usepackage{dcolumn}
\usepackage{bm}
\usepackage{theorem}
\usepackage{amsmath,amssymb}

\newcommand{\qed}{\nobreak \ifvmode \relax \else
\ifdim\lastskip<1.5em \hskip-\lastskip \hskip1.5em plus0em
minus0.5em \fi \nobreak \vrule height0.75em width0.5em
depth0.25em\fi}

\input{epsf}

\newcommand{\beq}{\begin{equation}}
\newcommand{\eneq}{\end{equation}}

\begin{document}
\title{Non-equilibrium entanglement in a driven Dicke model}
\author{Victor M Bastidas$^{1,2}$, John H Reina$^1$ and Tobias  Brandes$^2$ }
\address{$^1$Universidad del Valle, Departamento de F\'isica, A. A. 25360, Cali, Colombia}
\address{$^2$Institut f\"ur Theoretische Physik, Technische Universit\"at Berlin, Hardenbergstr. 36, 10623 Berlin, Germany}
\ead{vicmabas@univalle.edu.co and jhreina@univalle.edu.co}
\begin{abstract}
We study the entanglement dynamics in the externally-driven single-mode Dicke model in the thermodynamic limit, when the field is in resonance with the atoms. We compute the correlations in the atoms-field ground state by means of  the density operator that  represents the pure state of the universe and the reduced density operator for the atoms, which results from taking the partial trace over the field coordinates. As a measure of bipartite entanglement, we calculate the linear entropy, from which we analyze the entanglement dynamics. In particular, we found a strong relation between the stability of the dynamical parameters and the reported entanglement.
\end{abstract}
\noindent
To appear  in {\jpcs}

\section{Introduction}
Understanding the properties displayed by the entanglement of physical systems is one of the fundamental purposes of quantum information theory \cite{Nielsen}. A quantum system composed of two or more entangled subsystems has the interesting property that although the state of the total system can be well defined, it is impossible to identify individual properties for each one of its parts. A subject that has attracted recent interest is linked to the relationship between entanglement and certain physical properties of many body quantum systems; this has taken special interest in relation to quantum phase transitions (QPTs)  \cite{Legeza,Dusuel,Lidar}. A QPT in a many body system strongly influences the behaviour of the system near to the
critical point, with the consequent appearance of long-range correlations in the ground state. Although several proposals exist, currently there is no complete theory of multipartite entanglement, and the common  techniques are based on bipartite decompositions of the total system. This type of decomposition has made it possible to study the entanglement in the Dicke model in thermal equilibrium \cite{brandes1,brandes2,tolkunov}. Here, we study the entanglement properties of a many-body  system in the non-equilibrium regime, in which the total system has a unitary dynamics but is not an isolated system. For this purpose we study the single mode ``externally-driven" Dicke Hamiltonian
\begin{equation}
\label{eq:hamdN} 
\hat{H}(t) = \omega_{0}J_{z}+\omega
a^{\dagger}a+\frac{g(t)}{\sqrt{N}}(J_{+}+J_{-})(a^{\dagger}+a),
\end{equation}
where $J_z=\sum_{i=1}^{N}J^{(i)}_{z}$,  
$J_\pm=\sum_{i=1}^{N}J^{(i)}_{\pm}$ are collective atomic operators, $\omega_{0}$ is the level splitting of the atoms, $\omega$ is the frecuency of the bosonic mode
and $g(t)= g+\Delta g \cos \Omega
t$ is the time dependent atom-field coupling. We assume that $\Delta g$ is a fraction of the static coupling $g$. An exact diagonalisation of the problem Eq. \eqref{eq:hamdN} has been previously carried out by means of the Holstein-Primakoff
transformation for  the case of a {\it static} atom-field coupling $g(t)\equiv
g=constant$ \cite{brandes1,brandes2}, but there is no known  solution for the  time dependent atom-field coupling case. 

\section{The reduced density operator}
When studying a composed quantum system whose dynamics is unitary, it is interesting to study the dynamics of the parts of the system. In contrast to the dynamics of the whole system, the dynamics of the subsystems is not unitary. However, it is possible to make a description of the subsystem through the reduced density operator. The Schr\"odinger equation for the single mode ``externally-driven" Dicke Hamiltonian has an exact solution when the field is in resonance with the atoms $\omega =\omega_{0}$.
With the purpose of formulating the exact solution, we introduce an abstract coordinate representation through the coordinates of the physical field $x$ and the atoms coordinate $y$.
In terms of these coordinates, the ground state of the universe (atoms+field) is given by the expression \cite{vj}
\begin{equation}
\label{eq:solcs}
\Psi_{0_{-},0_{+}}(x,y,t)=
\exp(i\gamma_{0_{-}}(t))\exp(i\gamma_{0_{+}}(t)) 
\Phi_{0_{-}}^{-}\left(\frac{x}{\sqrt{2}}-\frac{y}{\sqrt{2}},t\right)
 \Phi_{0_{+}}^{+}\left(\frac{x}{\sqrt{2}}+\frac{y}{\sqrt{2}},t\right),
\end{equation}
where 
\begin{equation}
\label{eq:solgs} \Phi_{0_{\mp}}^{\mp}(w,t)= \left(\frac{1}{2\pi
|B^{\mp}(t)|^{2}}\right)^{1/4}\exp\left(i\frac{\dot{B}^{\mp}(t)}{2
B^{\mp}(t)} w^{2}\right), \nonumber
\end{equation}
and the time dependent phase is given by  
\begin{equation}
\label{phase}
\gamma_{0_{\mp}}(t) =
\int_{0}^{t} \langle \Phi_{0_{\mp}}^{\mp},t| i \frac{\partial}{\partial
t}-\hat{H}(t)|\Phi_{0_{\mp}}^{\mp},t \rangle d t. \nonumber
\end{equation}
It is interesting to note that the dynamics of the system is influenced by the dynamics of the auxiliary parameters $B^{\mp}(t)$, which are solutions of the Mathieu equation \cite{abram}
\begin{equation}
\label{eq:eqmath}
 \ddot{B}^{\mp}(t)+[(1 \mp 2g)\mp(2\Delta g)\cos
\Omega t]B^{\mp}(t)=0,
\end{equation}
subject to the Wronskian condition 
\begin{equation}
\label{eq:eqwr}
\dot{B}^{\mp}(t)(B^{\mp})^{\ast}(t)-B^{\mp}(t)(\dot{B}^{\mp})^{\ast}(t)=i. \nonumber
\end{equation}
We describe the dynamics of the total system in terms of the density operator, where the pure state of the universe is represented by the operator $\hat{\rho}_{G}(t)= |\Psi_{0_{-},0_{+}},t \rangle  \langle \Psi_{0_{-},0_{+}},t|$. In the coordinate representation, the density matrix takes the form
\begin{equation}
\label{eq:opdentot}
\rho_{G}(x',y';x,y,t)=\Psi ^{\ast}_{0_{-},0_{+}}(x,y,t)\Psi_{0_{-},0_{+}}(x',y',t).
\end{equation}
We establish a bipartite decomposition of the universe and study 
the reduced dynamics of the atoms through their reduced density matrix
(RDM). In so doing, we calculate the partial trace over the physical field coordinate $x$,
\begin{equation}
\label{eq:opdenred}
\rho^{(R)}_{G}(y',y,t)=\int_{-\infty}^{+\infty} \Psi ^{\ast}_{0_{-},0_{+}}(x,y,t)\Psi_{0_{-},0_{+}}(x,y',t) \ dx.
\end{equation}
The ground state of the universe $\Psi_{0_{-},0_{+}}(x,y,t)$ given in Eq. \eqref{eq:solcs} is Gaussian, which facilitates the integration of Eq. \eqref {eq:opdenred}. After some algebraic calculations, we obtain the following result for the reduced density matrix 
\begin{eqnarray}
\nonumber
\rho^{(R)}_{G}(y',y,t)&=&\left(\frac{\Re e (\xi^{-})\Re e
(\xi^{+})}{\pi(\Re e (\xi^{-})c^{2}+\Re e
(\xi^{+})s^{2})}\right)^{1/2}  
\,  \exp\left(\frac{c^{2}s^{2}[y(\xi^{-}-\xi^{+})^{\ast}+y'(\xi^{-}-\xi^{+})]^{2}}{4(\Re
e (\xi^{-})c^{2}+\Re e (\xi^{+})s^{2})}\right)\times\nonumber
\\&&\,\exp\left(-\frac{y^{2}}{2}[(\xi^{-})^{\ast}s^{2}+(\xi^{+})^{\ast}c^{2}]\right) \exp\left(-\frac{(y')^{2}}{2}[(\xi^{-})s^{2}+(\xi^{+})c^{2}]\right),
 \label{rd1}
\end{eqnarray}
where $c=s=1/\sqrt{2}$  and  $\Re e (\xi^{\mp})$ is the real part of the function
$
\xi ^{\mp}(t)=-i\frac{\dot{B}^{\mp}(t)}{B^{\mp}(t)}
$.
%
%
In the particular case $\Delta g = 0$ ($g(t)=g$), and $\omega=\omega_{0}$, this function is independent of time and becomes $\xi ^{\mp}(t) = \epsilon^{\mp}=\sqrt{(1 \mp 2g)}$. Hence, our model reproduces the result previously reported in the literature for the atom's reduced density matrix for the single mode Dicke model in thermal equilibrium \cite{brandes1,brandes2}.

\section{The linear entropy}
We consider a pure state of a composed system  (the universe), say system $AB$, represented by the density operator $\hat{\rho}_{AB}$. The linear entropy for the reduced density operator $\hat{\rho}_{A}=tr_{B}(\hat{\rho}_{AB})$ of subsystem A is defined by $L_{A}=1-tr_{A}(\hat{\rho}^{2}_{A})$, where $tr_{A}$ ($tr_{B}$)
denotes the trace over the subsystem A (B) \cite{Plenio07}.
\begin{figure*}
\label{figu1}
\begin{center}
\hspace{-1cm}
\includegraphics [scale=0.49]{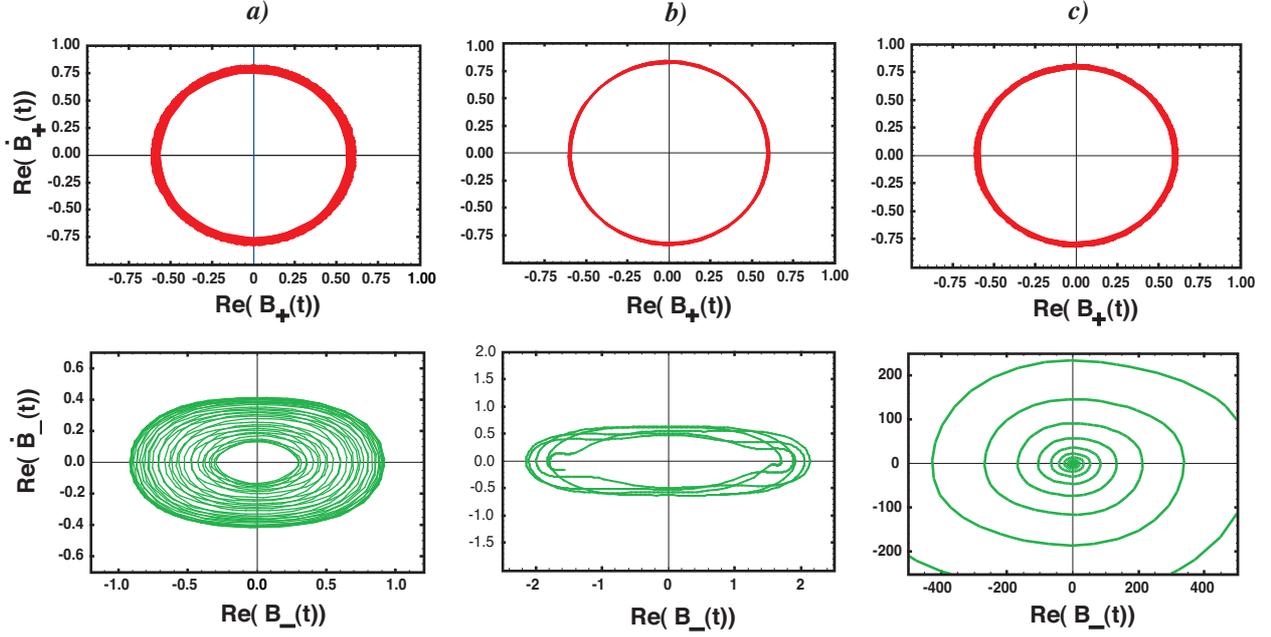}
\caption{Phase space trajectories of the auxiliary dynamical parameters $B^{\pm}(t)$ for $\Delta g =0.1 g$, and a value of the static coupling in a) the stable region $g=0.40$, b) the stable region $g=0.46$, and  c)  the unstable region $g=0.38$.}
\end{center}
\end{figure*}
This gives a measure of {\it purity} of the reduced density operator $\hat{\rho}_{A}$ (o  $\hat{\rho}_{B}$) of one part of the total system (in the bipartite decomposition of the universe).
If the pure state of the universe is separable, the reduced density operator of one part of the system represents a pure state and, as a result, its linear entropy must be zero. Similarly, if the pure state of the universe is a maximally entangled state \cite{Nielsen}, the linear entropy equals  $1/2$.
In the context of our  externally driven single mode Dicke model, it is possible to use the linear entropy as a measure of bipartite entanglement for the atoms-field system. The linear entropy for the atoms'  time dependent reduced density operator, Eq. \eqref{rd1}, is given by 
$
L(t)=1-tr\left[\left(\hat{\rho}^{(R)}_{G}(t)\right)^{2}\right].
$
By using the density matrix representation of Eq. \eqref{rd1}, we explicitly calculate the linear entropy as
\begin{equation}
L(t)=1-\int_{-\infty}^{+\infty}\int_{-\infty}^{+\infty}\rho^{(R)}_{G}(y',y,t)\rho^{(R)}_{G}(y,y',t)
\ dy \ dy', \nonumber
\end{equation}
which gives, after some algebraic calculations, the result 
\begin{equation}
L(t,g)=1-\frac{\pi \Lambda^{2}}{\sqrt{4(\Re e
(\alpha))^{2}-\beta^{2}}} \ ,
\label{entr1}
\end{equation}
where the parameters $\alpha$, $\beta$ and $\Lambda$ are defined as
\begin{eqnarray}
\nonumber
\alpha &=&
\frac{(\xi^{-})^{\ast}s^{2}+(\xi^{+})^{\ast}c^{2}}{2}-\frac{c^{2}s^{2}[(\xi^{-}-\xi^{+})^{\ast}]^{2}}{4(\Re
e (\xi^{-})c^{2}+\Re e (\xi^{+})s^{2})}  ,   \; 
\beta= \frac{c^{2}s^{2}(\xi^{-}-\xi^{+})^{\ast}(\xi^{-}-\xi^{+})}{2(\Re
e (\xi^{-})c^{2}+\Re e (\xi^{+})s^{2})} ,\\
 \Lambda&=&\left(\frac{\Re e (\xi^{-})\Re e
(\xi^{+})}{\pi(\Re e (\xi^{-})c^{2}+\Re e
(\xi^{+})s^{2})}\right)^{1/2}. 
\end{eqnarray}
\begin{figure*}
\begin{center}
\label{figu2}
\hspace{-1.0cm}
\includegraphics[scale=0.31]{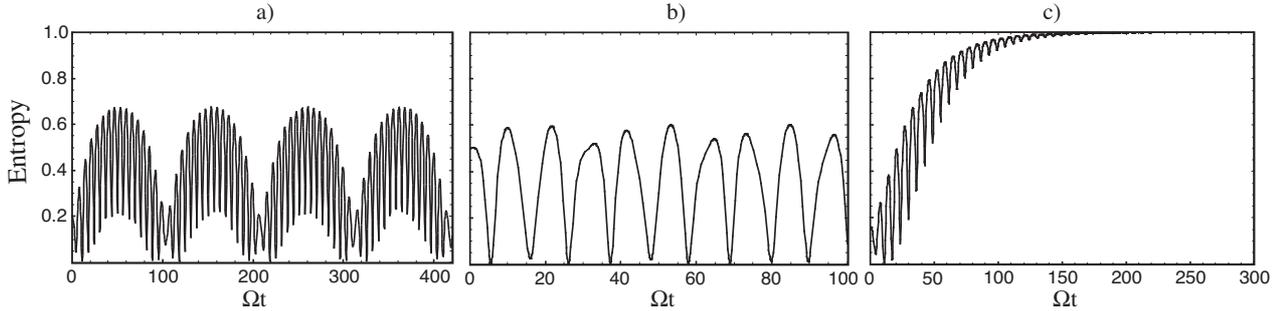}
\caption{Linear entropy as a function of time
for periodic atom-field coupling with $\Delta g =0.1 g$; a) $g=0.40$, b) $g=0.46$, and c) $g=0.38$.}
\end{center}
\end{figure*}

\section{Results}
Interestingly, the linear entropy exhibits a time dependence that is not determined by the global phase of the ground state wave function of the universe, Eq. \eqref{eq:solcs}, but by the auxiliary dynamical parameters $B^{\mp}(t)$.
In the particular case $\Delta g =0.1 g$, and $\Omega = 1$, the stability zones of Eq. \eqref{eq:eqmath} are known \cite{vj}, and the study of the corresponding solutions is based on the Floquet's theorem for second order
differential equations with time periodic coefficients. For this, the solutions of  Eq. \eqref{eq:eqmath} have the general form $B^{\pm}(t)= \exp (i F^{\pm} t)\phi^{\pm}(t),$
where $\phi^{\pm}(t+T)=\phi^{\pm}(t)$ and $F^{\pm}$ is the Floquet
exponent, which  depends on the shape of the driving.  For
driving functions for which $F^{\pm}$ is complex,  the
solution becomes unstable. In the stable regime, $F^{\pm}$ is a real
number \cite{abram}. In Figs. 1(a), (b) we consider the phase space representation of the trajectories of the auxiliary dynamical parameters $B^{\pm}(t)$. For values of $g$ that belong to a common stability zone ($g=0.4$ and $g=0.46$) \cite{vj} the solutions are bounded in the phase space. In order to establish a relationship between the stability and the entanglement dynamics, we consider the linear entropy in Figs. 2(a), and (b) for the values of the static coupling described above. Our choice of the initial
conditions for Eq. \eqref{eq:eqmath}  implies that the system starts in thermal equilibrium,
but is described by a non-separable quantum state. Strikingly, the dynamics exhibits a behaviour whereby the system experiences a disentanglement process, before reaching a maximum value of entanglement, and then successive collapses and revivals.

Figure 1(c) shows another interesting behaviour: when the static coupling belongs to the instability zone ($g=0.38$) of $B^{-}(t)$, the solution $B^{+}(t)$ is bounded while the solution $B^{-}$ is unbounded in the phase space. For this value of the static coupling, the linear entropy oscillates, before it reaches the stationary state with linear entropy $L=1$, as shown in Fig. 2(c).
In \cite{vj} we discuss the relationship between the stability zones and the localization of the ground state wave function of the universe. In order to study this relation, we define the characteristic length $l^{\mp}(t)= \sqrt{2} |B\mp (t)|$, which  is bounded in the stable zones and unbounded in the unstable zones. The numerical simulations of the ground state probability density show that for values of the static coupling that belong to the common stability zones the probability is localized in the abstract $x-y$ space and presents an oscillatory behaviour. In contrast, for values of the static coupling that belong to the unstable zones, the probability density has an oscillatory behaviour. However, when the system evolves, the probability density is systematically stretched in a fixed direction, with a consequent dilation in the perpendicular direction. 

\section{Conclusions}
We have obtained exact results for the atoms' reduced density operator and the linear entropy for the externally driven Dicke model, in the thermodynamic limit. These results allow us to study the entanglement dynamics through the auxiliary dynamical parameters. In contrast, in the Dicke model in thermal equilibrium \cite{brandes1,brandes2},  the atoms-field entanglement does not have dynamics because the temporal dependence of the ground state of the universe is given by a global phase of the wave function.
In this work, the auxiliary dynamical parameters and its stability properties determine the entanglement dynamics of the system and the behaviour of the universe ground state wave packet.
\ack 
We acknowledge financial support from Colciencias  under
contract 1106-452-21296, and the scientific exchange program PROCOL (DAAD-Colciencias).

\end{document}